\documentclass[11pt]{article}

\usepackage{epsfig}

\begin{document}

\title{
{\bf Another view on the velocity at the Schwarzschild horizon}}

\author{ Ismael Tereno \\
{\em Centro de Fisica Nuclear, Universidade de Lisboa} \\
{\em 1649-003 Lisboa, Portugal}
}
\date{}
\maketitle

\begin{abstract}
It is shown that a timelike radial geodesic does not become null at
the event horizon.
\end{abstract}

Recently an attempt was made to demonstrate that in the Schwarzschild
geometry the radial geodesics of material particles become null at the
event horizon \cite{mitra1}.

For this purpose, was derived an expression that corresponds to the
velocity of a material particle following a radial trajectory as measured
by an observer, also on a radial trajectory, when they intersect.
The observer mantains its spacelike Kruskal coordinates unchanged and for
this reason we call it a Kruskal observer. For $r>2m$, the expression is,
(eq.(20) of \cite{mitra1} and eq.(8) of \cite{tereno}),

\begin{equation}
\label{v}
v={{1+\tanh(t/4m){{dt}\over{dr}}(1-2m/r)}
\over{\tanh(t/4m)+{{dt}\over{dr}}(1-2m/r)}},
\end{equation}
where $dt$ and $dr$ refer to the movement $t(r)$ of the particle.
At the event horizon, where $r=2m$ and $t=+\infty$, the value of
eq.(\ref{v}) is apparentely indetermined and in \cite{mitra1} it is stated
that $v=1$, independently of the precise relationship $t(r)$.

In \cite{tereno} some manipulations were made maintaining the generality
of the expression, i.e. without substituting for $t(r)$. These permited to
show (eq.(13) of \cite{tereno}) that the velocity is always less than 1
along the way, until it obviously turns to $0/0$ at $r=2m$. So there is no
a priori reason to think it is necessarily $v=1$.

This procedure was commented in a somewhat ungracious manner in
\cite{mitra2} without any further explanations being made.

The best way to avoid confusion and get a definitive result seems to be
to consider a specific geodesic $t(r)$, transforming (\ref{v}) in a
function of 1 variable.

Let us then consider a material particle in an ingoing radial geodesic
parametrized by its proper time $\tau$. For this trajectory we can write,
in Schwarzschild coordinates, 

\begin{equation}
ds^2=-d\tau ^2 = -\left( {1-{{2m} \over r}} \right)dt^2+\left( {1-{{2m}
\over r}} \right)^{-1}dr^2.
\label{metschw}
\end{equation}
Inserting the conserved quantity for motion,

\begin{equation}
E={{dt} \over {d \tau}} \left(1-{{2m} \over {r}}\right),
\end{equation}
we obtain,

\begin{equation}
\label{dtdr}
{{dt} \over {dr}}=-E\left(1-{{2m} \over {r}}\right)^{-1}
\left[ E^2-\left(1-{{2m} \over {r}}\right) \right]^{-1/2}.
\end{equation}
Each geodesic is characterized by its value of $E$, defined by the
initial conditions $r_i$ and $v_i$ ( $v_i$ is refered to the observer
at rest at $r_i$).

\begin{equation}
\label{condit}
E=\left(1-{{2m} \over {r_i}}\right)(1-v_i^2)^{-1}.
\end{equation}

To simplify the integration of $dt/dr$ we introduce the approximation,

\begin{equation}
\label{approx}
\left(1-{{1-2m/r}\over{E^2}} \right)^{-1/2} \approx
1+{{1-2m/r}\over{2E^2}},
\end{equation}
valid for small $r-2m$.

This way we obtain,

\begin{equation}
\label{t}
t(r)=t_0+\int\limits_{r}^{r_0} {{s \over {s-2m}}ds+{1 \over {2E^2}}}
(r_0-r)=t_0+(r_0-r)\left( {1+{1\over {2E^2}}} \right)+2m\ln \left|
{{{r_0-2m} \over {r-2m}}} \right|.
\end{equation}
Each geodesic $E$ begins at a point $r_i$ at a time $t_i$ which we may
consider to be $t_i<0$. While it does not reach $r_0$ we do not know the
exact expression for $t(r)$. $r_0$ is the point where the approximation
(\ref{approx}) becomes valid and we make $t_0=0$.

Now we can insert (\ref{t}) and (\ref{dtdr}) in (\ref{v}) and plot
$v(r)$ for several geodesics $E$.

\begin{figure}[ht]
\label{fig:figure1}
\begin{center}
\epsfig{file=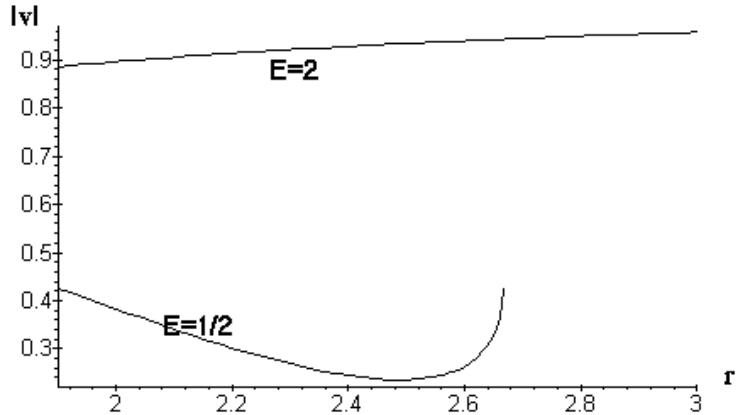, height=6cm}
\caption{Velocities at the region of the horizon.}
\end{center}
\end{figure}

In figure 1 we plot $|v|$ for two values of $E$ assuming
$r_0=3m$. Each curve represents the velocity of one particle measured
by a family of Kruskal observers, each one intersecting the particle
at a different point between $3m$ and $2m$. Even though eq. (\ref{v})
was defined as a velocity only for $r>2m$, we plot that mathematical
function through $r=1.9m$ to get a clearer view that $v(r=2m)$ is
less than 1.

Using (\ref{condit}) we note that the farthest point $r_i$ where a
geodesic with $E<1$ can begin is $r_i=2m/(1-E^2)$. For $E=1/2$ we
get $r_i=(8/3)m$ and that is why that curve in the figure does not reach
$r=3m$.

We can also get valuable information from the plot of $v$ at $r=2m$ as
a function of $E$.

\begin{figure}[ht]
\begin{center}
\epsfig{file=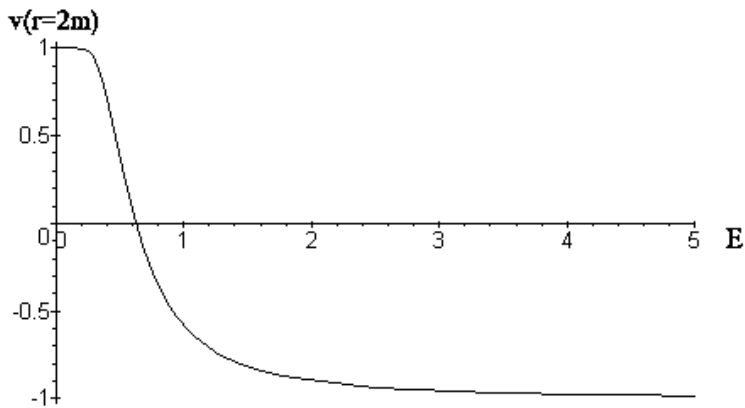, height=6cm}
\caption{Velocities at the horizon.}
\end{center}
\label{fig:figure2}
\end{figure}

In figure 2 we can see that $|v|$ tends to 1 only at the
limits of the domain of $E$. That is when $E=+\infty$ which is a null
geodesic and when $E=0$. From eq.(\ref{condit}) we see the latter
corresponds to a geodesic with $r_i=2m$ which is at rest relatively
to the horizon. This corresponds to the well known result that a
particle at rest on the horizon must be a photon and its
velocity is 1 relatively to a radial observer.

For all the other cases the modulus of the velocity plotted in
figure 2 is less than 1.

For greater values of $E$, $v$ is negative which means the particle
follows the observer and reaches it from one side. For smaller
values of $E$, $v$ is positive which means the observer follows the
particle and see it approaching from the other side of the $r$ axis.

This result is in agreement with the one presented in \cite{janis}. There
a new set of coordinates is introduced. The transformations are,
essentially,

\begin{equation}
\label{coojanis}
\left\{
\begin{array}{lll}
\mbox{$x_0=(w-r)/\sqrt{2}$} \\ \\
\mbox{$x_1=(w+r)/\sqrt{2}$} \\ \\
\mbox{$x_2=2m\theta$}\\ \\
\mbox{$x_3=2m\varphi$}
\end{array}
\right.
\end{equation}
where $w$ is the ingoing Eddington-Finkelstein coordinate,

\begin{equation}
\label{vedfink}
w(t,r)=t+r+2m\ln \left| {{{r-2m} \over {2m}}} \right|.
\end{equation}

In these coordinates the metric takes the form,

$$
ds^2=\left[ {-{1 \over 2}\left( {1-{{2\sqrt {2} m} \over {x_1-x_0}}}
\right)
-1} \right] dx_0^2 - \left( {1-{{2\sqrt{2} m} \over {x_1-x_0}}} \right)
dx_0dx_1+\left[ {-{1 \over 2} \left( {1-{{2 \sqrt{2} m} \over {x_1-x_0}}}
\right)+1} \right]dx_1^2+ $$

\begin{equation}
+\left( {{{x_1-x_0} \over {2 \sqrt{2} m}}} \right)^2\left[ {dx_2^2+\sin ^2
({{x_2} \over {2m}})dx_3^2} \right],
\end{equation}
which at the horizon $(x_1-x_0=2 \sqrt{2})$ (and with $\theta=\pi /2$)
reduces to the Minkowski form,

\begin{equation}
\label{Minkow}
ds^2=-dx_0^2+dx_1^2+dx_2^2+dx_3^2.
\end{equation}

Let us now define the Janis observer as the one who mantains the 
spacelike coordinates $x_1, x_2, x_3$ constant. Like the Kruskal
observer it follows a radial geodesic. The velocity of a material
particle that moves along an ingoing geodesic $dt/dr$ as measured
by this observer is,

\begin{equation}
\label{vjanis}
v_1={{dx_1} \over {dx_0}}=\left( {{{1+dr/dt{r \over
{r-2m}}+dr/dt} \over {1+dr/dt{r \over {r-2m}}-dr/dt}}} \right).
\end{equation}

This expression is written as a function of only 1 variable $(r)$.
Inserting $dr/dt$ from (\ref{dtdr}) and approximating the squared factor
in that expression analogously to what was made in eq.(\ref{approx}),
we obtain,

\begin{equation}
\label{vfjanis}
v_1^2=\left( {{{{1 \over {2E}}-E+{{1-{{2m} \over r}} \over {2E}}}
\over {{1 \over {2E}}+E-{{1-{{2m} \over r}} \over {2E}}}}} \right)^2.
\end{equation}
For $r=2m$ we get,

\begin{equation}
v_1^2 (r=2m)= \left({{{1-2E^2} \over {1+2E^2}}} \right)^2,
\end{equation}
which is eq.(10) of \cite{janis}.

From here we see that $v_1<1$ unless $E=0$ or $E=+\infty$. In fact
the graph of this function $v_1(E)$ is identical to the one in
figure 2.

This is the general behaviour for the relative velocity of two moving
material particles at the Schwarzschild horizon.


\bigskip \bigskip

\begin{thebibliography}{99}

\bibitem{mitra1}
Mitra,A. (1999), astro-ph/9904162

\bibitem{tereno}
Tereno,I. (1999), astro-ph/9905144

\bibitem{mitra2}
Mitra,A. (1999), astro-ph/9905175

\bibitem{janis}
Janis,A. (1973) \emph{Phys.Rev.D} {\bf 8}, 2360

\end{thebibliography}
\end{document}